\documentclass[]{mn2e}
\usepackage{epsfig}
\title{On the current status of open-cluster parameters}
\author[E.~Paunzen, M.~Netopil]
{E.~Paunzen, M.~Netopil \\
Institut f{\"u}r Astronomie der Universit{\"a}t Wien, T{\"u}rkenschanzstr.
17, 1180 Wien, Austria \\
}
\pubyear{2006}
\begin{document}
\maketitle
\begin{abstract}
We aim to characterize the current status of knowledge on the accuracy
of open-cluster parameters such as the age, reddening and distance.
These astrophysical quantities are often used to study the global
characteristics of the Milky Way down to very local stellar phenomena.
In general, the errors of these quantities are neglected or
set to some kind of heuristic standard value. We attempt to give some realistic estimates for the accuracy of available cluster parameters by
using the independent derived values published in the literature. In total,
6437 individual estimates for 395 open clusters were used in our statistical
analysis. We discuss the error sources depending on theoretical as
well as observational methods and compare
our results with those parameters listed in the widely used catalogue by Dias
et al. (2002). In addition, we establish a list of 72 open clusters with the
most accurate known parameters which should serve as a standard table 
in the future for testing isochrones and stellar models.  
\end{abstract}
\section{Introduction}

The study of open clusters naturally introduces many advances,
because they are physically related groups of stars held together
by mutual gravitational attraction that were formed at roughly
the same time from one large cosmic gas and dust cloud. Their
evolutionary stages range from clouds where star formation still takes  
place at this moment to very old aggregates with turn-off points
as late as solar type stars. Therefore,
they represent samples of Population I stars of constant age and comparable 
intrinsic
chemical composition, best suited to study processes related to 
stellar evolution and formation, and to fix lines or loci in
several most important astrophysical diagrams. These procedures
are statistical methods, for example fitting isochrones to determine
the age, reddening, and distance of an open cluster, independent of
individual peculiarities of members. 

As the open clusters drift along their orbits, they are excellent tracers
for the global kinematics and dynamical characteristics of the Milky Way itself.
However,
some of their members escape the cluster, due to velocity changes in mutual 
close encounters or tidal forces in the galactic 
gravitational field. The escaped stars continue to orbit  
on their own as field stars.

\begin{figure}
\epsfxsize = 80mm
\epsffile{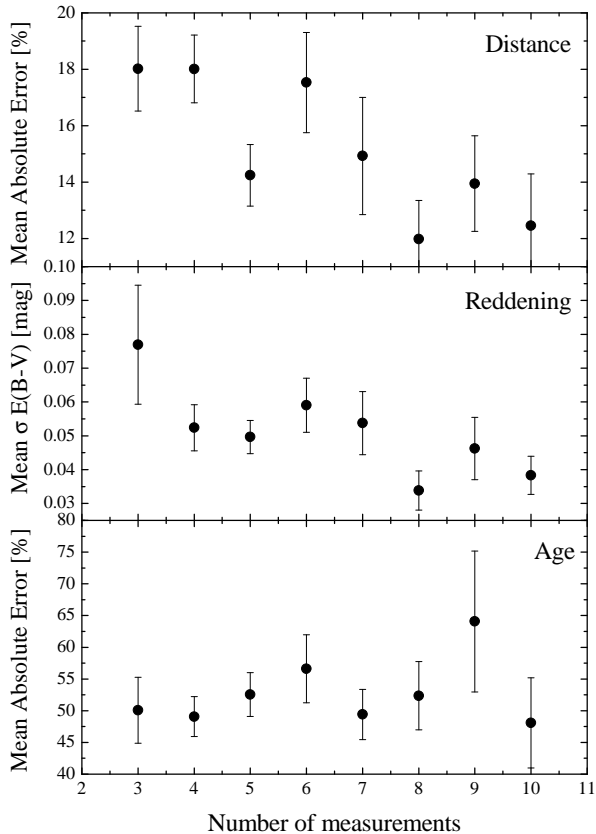}
\caption{The errors of the mean for the distance, reddening and age 
versus the number of available estimates in the literature for the 
complete sample of 395 open clusters. They are smaller for the distance and reddening
if more observations are available, but the mean errors for the age
estimate are more or less constant. There are only few open clusters with more
than 10 measurements which are not included in these diagrams to avoid a bias due to
poor number statistics.}
\label{means}
\end{figure} 

As a summary it can be concluded that open clusters allow the study of a significant
number of astrophysical parameters and models from the global scale of the Milky
Way down to processes in individual stars from the Pre-Main-Sequence to Post-AGB
evolution.

But what about the accuracy of the presently known open-cluster parameters? 
Surprisingly enough, there is, to our knowledge, no work within the last ten years
which deals with this question in detail. Although there are several new catalogues
(Dias et al. 2002 and Kharchenko et al. 2005) published and a flood of new parameter 
estimates, mainly because the common use of CCDs permits the
observation of fainter and more distant
clusters, there is no published statistical analysis describing if and how the accuracy
has improved.

If we look at recent statistical papers dealing with open-cluster parameters
(e.g. Chen, Hou \& Wang 2003, Dias \& L{\'e}pine 2005), none
of them have any detailed error treatment included. Although apparent ``mean values''
for the age, distance and metallicity are used, the level of accuracy is completely
ignored in the final conclusions. 

In this work we try to fill this gap by presenting a detailed analysis of the
current available open-cluster parameters and the corresponding errors. For this purpose
we have searched the literature up to November 2005 for individual cluster
parameter estimates. The primary goal
is to establish a list of ``standard'' open clusters covering a wide range of ages,
reddenings and distances from the Sun selected on the basis of the smallest errors
from the available parameters in the literature. These clusters should serve as a
standard set for future investigations. We also compare the derived mean 
values with those of the most recent catalogue by Dias et al. (2002) resulting in an 
overall satisfactory agreement.

\section{Data selection and preparation} \label{dsap}

For the determination of mean values for
the age, reddening and distance, we have used the following data
sources: Janes, Tilley \& Lyng\aa, (1988), Malysheva (1997), Dambis (1999)
Dutra \& Bica (2000), Loktin et al. (2001), Tadross (2001),
Lata et al. (2002, and references therein), Kharchenko et al. (2005) and
an updated list of individual publications as listed in
WEBDA (http://www.univie.ac.at/webda/recent\_data.html) which includes
more than 300 different papers. The latter was mainly performed within the
framework of ADS
(http://cdsads.u-strasbg.fr/). The individual references were further checked
for sources from the literature to guarantee a maximum of available
estimates accessible to the community. All selected publications list $E(B-V)$,
log\,$t$ and the distance from the Sun. The only exception is Malysheva (1997)
who lists $E(b-y)$ which was converted by a factor of 1.43 to $E(B-V)$. 
For the further analysis we have used $A_V$\,=\,3.1$E(B-V)$
which is the best mean value for most regions in the Milky Way (Winkler 1997).

We have carefully checked
whether the listed parameters for each open cluster in the individual 
references are independent of each other and not used by two authors
twice. Kharchenko et al. (2005), for example, used
data by Loktin et al. (2001). Such duplicated data were not used in the 
subsequent analysis.

Janes et al. (1988) studied all photometric
measurements available at that time to them
and introduced weights to derive the cluster parameters.
They summarize the knowledge and results for the given time and supersede the catalogue by Lyng\aa\, (1987).  

The catalogue of Dias et al. (2002) was not used at this stage because
it already lists averaged values from the
literature. Furthermore we made a comprehensive comparison 
with the results of this work (Sect. \ref{compd}).

The metallicity is also a free parameter when fitting isochrones
to photometric data. Chen et al. (2003) compiled a catalogue
of metallicities and analysed this sample. They derived a radial
iron gradient of $-$0.063(8)\,dex\,kpc$^{-1}$ for the Milky Way with an
intrinsic spread of the overall open cluster metallicities of $\pm$0.2\,dex.
This is in excellent agreement with the scatter of the metallicity
found for F and G type stars in the solar neighbourhood (Karatas, Bilir 
\& Schuster 2005). The effect of different metallicities to the isochrone
fitting procedure is extensively discussed by Pinsonneault et al. (2004).
The $\Delta M_V$ values range from +0.1 to $-$0.3\,mag for [Fe/H]\,=\,$\pm$0.2\,dex
and constant effective temperature sensitive color indices. Taking a difference
of 0.3\,mag for the distance modulus introduced by an incorrect treatment of
the metallicity, the error for 1\,kpc is 150\,pc or 15\%. However, these 
discrepancies are only valid for effective temperatures cooler than 8000\,K.
For hotter temperatures, the difference in the metallicities result 
mainly in a shift of the temperature sensitive color indices (Schaerer et al.
1993). Lebreton (2000) presented a comparison of galactic field stars with
known metallicities as well as accurate parallax measurements and appropriate
isochrones (see Figure 6 therein). There is an apparent mismatch at any 
level of metallicity. However, the most significant differences are at
large underabundances ($>$\,$-$2.5\,dex) which are normally not found in
open clusters of the Milky Way. There are only a few independent estimates
of the metallicity in the literature. Twarog, Ashman \& Anthony-Twarog (1997),
Gratton (2000) and Chen et al. (2003)
give a summary of the current knowledge with an error estimate of the
individual values. Since then very few new estimates have been published.
We have therefore not made a new and redundant error analysis of the
metallicities for open clusters.

All individual data were considered as completely equal and no weights
were attributed. No assessment of the corresponding reference for the
used methods, isochrones and so on was done. This should guarantee that no
unknown bias is introduced in the data set. We have rejected NGC~2451,  which recent studies have shown that is
formed by the superposition of two clusters at different distances, and Basel~11, 
which was wrongly assigned in the original survey. Therefore
the former parameters determined for the joined data are no longer significant.
Furthermore, Pena \& Peniche (1994) reported significant inconsistencies for the
analysis of NGC~7209 which we also neglected.

We have also not included the given ``intrinsic'' errors of individual
references. The simple reason is that most authors list only an overall
error (if any) for all measurements which is highly unrealistic and would introduce
an unknown bias.

For the final list, we have averaged the data of open clusters for which at
least three independent estimates of the age, reddening and distance are
available. It contains 6437 individual estimates for 395 open 
clusters. As the last step, the standard deviations were calculated.

\begin{figure}
\epsfxsize = 82mm
\epsffile{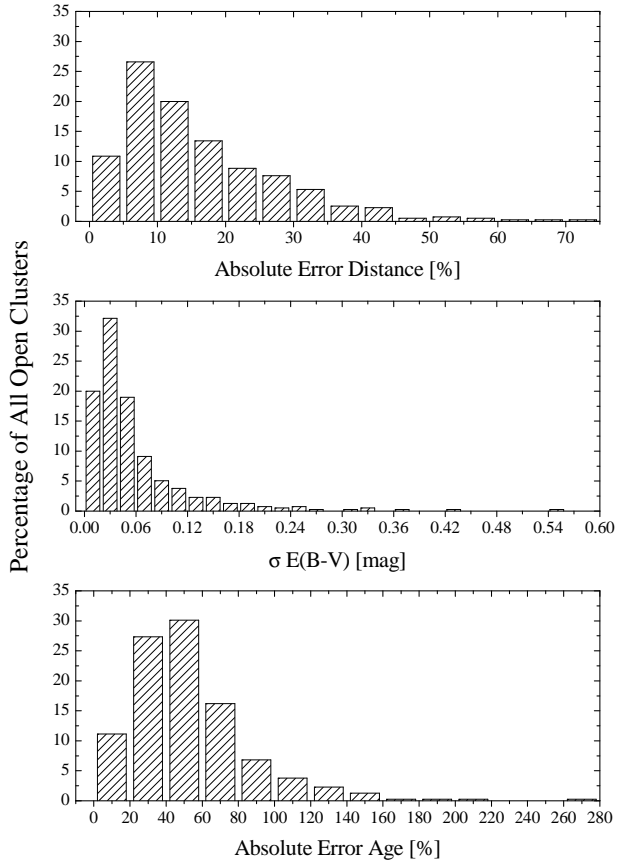}
\caption{The distribution of the mean absolute errors of the distance and age as
well as the standard deviation of the reddening for the complete
sample of 395 open clusters with more
than three independent measurements from the literature (Section \ref{analysis}).
This figure summarizes the current accuracy of open-cluster parameters.} 
\label{means_d}
\end{figure}

\begin{figure}
\epsfxsize = 82mm
\epsffile{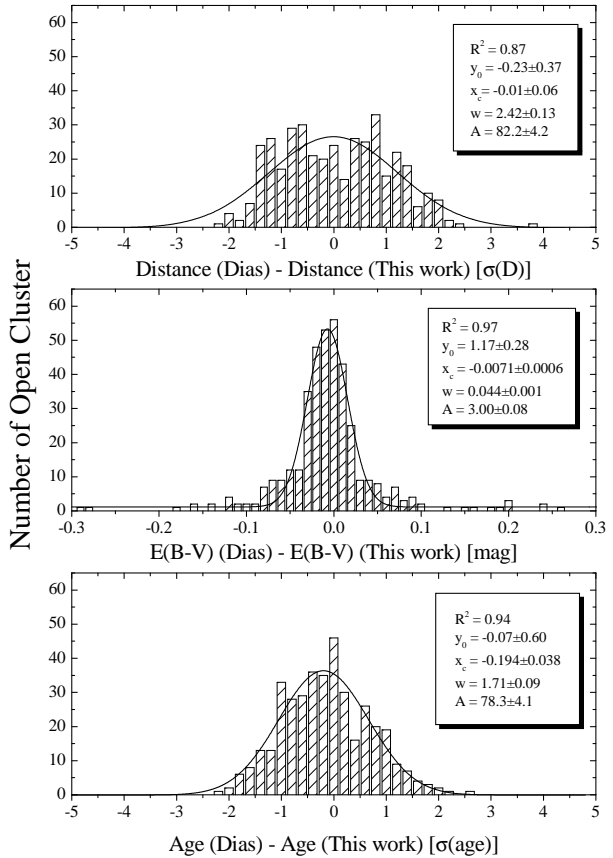}
\caption{Distribution of the differences of the distance, age (both in units of the
standard deviation of the mean) and reddening 
from the most recent catalogue by Dias et al. (2002) and the estimate
from this work. The values of the Gaussian fits are shown in the panels.} 
\label{dias}
\end{figure}

\section{Statistical analysis of the sample} \label{analysis}

Fitting isochrones to photometric data is not straightforward. 
The goodness of the fit depends on the following characteristics of the colour-magnitude
diagram of an open cluster:
\begin{itemize}
\item Number of members
\item Percentage of contamination by non-members
\item Age, i.e. visibility of the turn-off point, an important source
of errors for young open clusters
\item Presence of spectroscopic binaries which shifts the main sequence by
a maximum of 0.754\,mag 
\item Presence of a red giant clump which makes an isochrone fitting much
more significant
\item Differential reddening
\item Available photometric indices, $(U-B)$ is, for example, an excellent
indicator for the reddening and membership
\item Quality of the data
\item Metallicity
\end{itemize}
In addition, there is also a ``human factor'' which enters the estimate
of the best fit to observational data. Andersen \& Nordstr{\"o}m (1999) show how different research groups
interpret the same data set quite differently, using the excellent example of NGC 3680 (Figure 3 therein). This is not because one set
of parameters is better than the other but the technique of isochrone fitting
itself is not unambiguous. For this reason, very sophisticated computer algorithms
were developed to overcome this problem (J{\o}rgensen \& Lindegren 2005). The method
is rather simple: the locations of the color-magnitude diagrams are divided into boxes and the 
incidence of members within each box is counted. Those numbers are compared with
the same parameterization of standard isochrones. Loktin et al. (2001) and
Kharchenko et al. (2005), for example, used such a black box
to derive cluster parameters. However, J{\o}rgensen \& Lindegren (2005) concluded
that the systematic errors of computer algorithms are at least on the same
level as those introduced by the human factor.

Another error source is the choice of appropriate isochrones. Grocholski \& Sarajedini
(2003) discussed the apparent differences of the five most commonly used sets of
isochrones. They concluded that none of the theoretical isochrones reproduce
the observational data in a consistent manner over the
magnitude and colour range of the zero age main sequence.
In particular, there are significant zero point and shape differences
between the models and the observations. There are three reasons for these discrepancies:
1) the input physics has to be improved; 2) The transformation between theoretical
and observational parameters; 
and 3) more standard open clusters have to be established for testing new models.

For our statistical analysis we have chosen the following parameters:
\begin{itemize}
\item $\sigma(D)/D$: the absolute error of the mean distances
\item $\sigma\,E(B-V)$: the standard deviation of the mean reddening values
\item $\sigma(age)/age$: the absolute error of the mean ages (in units
of Myrs and not in the logarithmic scale) 
\end{itemize}
The parameters themselves are, in general, {\it not} independent from each
other. For example, the distance depends on the total absorption.  

Figure \ref{means} shows the errors of the mean for the 
distance, reddening and age 
versus the number of available estimates in the literature for the 
complete sample of 395 open clusters. 
The errors of the means for the distance and reddening decrease with more available observations while the mean errors for the age
estimate are more or less constant over the whole range. Note that there are also
a few parameters which are estimated more than ten times, but these few
data are not shown to avoid any poor number statistics.

The distribution of the different error estimates (Fig. \ref{means_d}) gives an
idea of the current accuracy for the open-cluster parameters. The reddening is
best known mainly because the main sequence can be fitted more easily for almost
all open clusters. For about 90\% of all clusters, the error is below 0.1\,mag.
The distances are also rather well determined: for about 80\% of all aggregates, the
absolute error is less than 20\%. As expected, the estimate of the age of
open clusters is the most difficult and introduces large uncertainties. Only
11\% of the investigated open clusters have errors less than 20\%. Even more than
30\% exhibit an absolute error larger than 50\% with extreme values of more than
200\%. These characteristics are most certainly not due to wrong estimates or
typographical errors in the reference but an intrinsic problem of the method itself.

How should one overcome the problem of a proper age estimate for open clusters? We
have shown that the knowledge of the reddening and determining distances is satisfactory. If we look at the metallicity, the situation is rather poor. Most
authors use solar abundances (e.g. Dutra \& Bica 2000, Loktin et al. 2001 and 
Kharchenko et al. 2005) throughout the isochrone fitting process, this procedure is in most cases justified, but normally photometric data will not allow us
do it any better. We believe that the only way out of the dilemma is to
establish ``standard'' open clusters covering a wide range of distances, ages and reddening values 
which should allow, for example the testing of isochrones.

\begin{table*}
\caption[]{List of 72 suggested standard open clusters, sorted after the galactic longitude, 
from the literature.
The listed parameters are calculated mean values with their standard
deviations and should be treated in a statistical sense. The columns ``$No$''
denote the number of used independent estimates.} 
\label{st_ocls}
\begin{center}
\begin{tabular}{llrrrrrccrrrr}
\hline
\multicolumn{1}{c}{Cluster} & \multicolumn{1}{c}{Name} &  \multicolumn{1}{c}{$l$} & \multicolumn{1}{c}{$b$} & 
\multicolumn{1}{c}{$D$} & \multicolumn{1}{c}{$\sigma(D)$} & \multicolumn{1}{c}{$No$} & \multicolumn{1}{c}{$E(B-V)$} &
\multicolumn{1}{c}{$\sigma\,E(B-V)$} & \multicolumn{1}{c}{$No$} & \multicolumn{1}{c}{$age$} & \multicolumn{1}{c}{$\sigma(age)$} & 
\multicolumn{1}{c}{$No$} \\
& & & & \multicolumn{1}{c}{[pc]} & \multicolumn{1}{c}{[pc]} & & \multicolumn{1}{c}{[mag]} & 
\multicolumn{1}{c}{[mag]} & & \multicolumn{1}{c}{[Myr]} & \multicolumn{1}{c}{[Myr]} \\
\hline
Collinder 394 & C1850$-$204 & 14.88 & $-$9.24 & 648 & 41 & 4 & 0.25 & 0.01 & 5 & 74 & 6 & 5 \\
Berkeley 81 & C1859$-$005 & 34.51 & $-$2.07 & 2798 & 508 & 4 & 1.00 & 0.01 & 4 & 1000 & 1 & 4 \\
IC 4756 & C1836+054 & 36.38 & +5.24 & 415 & 46 & 5 & 0.19 & 0.01 & 6 & 674 & 128 & 5 \\
NGC 6834 & C1950+292 & 65.70 & +1.19 & 2147 & 59 & 4 & 0.73 & 0.03 & 4 & 65 & 18 & 4 \\
NGC 6791 & C1919+377 & 69.96 & +10.90 & 4418 & 743 & 7 & 0.17 & 0.04 & 7 & 7850 & 2026 & 7 \\
NGC 6871 & C2004+356 & 72.64 & +2.05 & 1675 & 132 & 7 & 0.46 & 0.03 & 8 & 9 & 2 & 8 \\
NGC 6819 & C1939+400 & 73.98 & +8.48 & 2188 & 255 & 5 & 0.23 & 0.08 & 5 & 2172 & 442 & 5 \\
NGC 6811 & C1936+464 & 79.21 & +12.02 & 1146 & 112 & 5 & 0.14 & 0.03 & 6 & 635 & 138 & 6 \\
NGC 7044 & C2111+422 & 85.89 & $-$4.15 & 3097 & 145 & 6 & 0.63 & 0.06 & 6 & 1824 & 361 & 6 \\
NGC 6939 & C2030+604 & 95.90 & +12.30 & 1321 & 274 & 6 & 0.45 & 0.08 & 6 & 1639 & 452 & 6 \\
King 10 & C2252+589 & 108.48 & $-$0.40 & 3373 & 167 & 4 & 1.15 & 0.01 & 4 & 45 & 11 & 4 \\
NGC 7789 & C2354+564 & 115.53 & $-$5.39 & 1924 & 182 & 8 & 0.25 & 0.02 & 9 & 1519 & 202 & 8 \\
Berkeley 99 & C2319+714 & 115.95 & +10.11 & 4902 & 51 & 4 & 0.30 & 0.01 & 4 & 3162 & 1 & 4 \\
King 12 & C2350+616 & 116.12 & $-$0.13 & 2453 & 124 & 5 & 0.58 & 0.03 & 6 & 11 & 1 & 5 \\
Berkeley 2 & C0022+601 & 119.71 & $-$2.31 & 5129 & 383 & 4 & 0.80 & 0.01 & 4 & 794 & 1 & 4 \\
NGC 129 & C0027+599 & 120.27 & $-$2.54 & 1632 & 56 & 8 & 0.56 & 0.03 & 9 & 62 & 15 & 9 \\
Trumpler 1 & C0132+610 & 128.22 & $-$1.13 & 2356 & 511 & 9 & 0.57 & 0.04 & 9 & 30 & 6 & 9 \\
Berkeley 64 & C0217+656 & 131.92 & +4.61 & 3860 & 229 & 5 & 1.05 & 0.01 & 5 & 1000 & 0.01 & 5 \\
NGC 744 & C0155+552 & 132.40 & $-$6.16 & 1214 & 176 & 3 & 0.41 & 0.05 & 5 & 184 & 49 & 5 \\
Melotte 20 & Alpha Per & 146.57 & $-$5.86 & 173 & 12 & 8 & 0.09 & 0.03 & 9 & 43 & 18 & 9 \\
NGC 1245 & C0311+470 & 146.65 & $-$8.93 & 2593 & 299 & 9 & 0.27 & 0.03 & 10 & 1065 & 276 & 10 \\
King 7 & C0355+516 & 149.77 & $-$1.02 & 2211 & 200 & 5 & 1.27 & 0.05 & 5 & 603 & 121 & 5 \\
NGC 1798 & C0508+475 & 160.70 & +4.85 & 4160 & 232 & 4 & 0.51 & 0.01 & 4 & 1421 & 16 & 4 \\
Berkeley 18 & C0518+453 & 163.63 & +5.02 & 4772 & 723 & 4 & 0.47 & 0.01 & 4 & 4271 & 241 & 4 \\
Melotte 22 & Pleiades & 166.57 & $-$23.52 & 133 & 9 & 8 & 0.05 & 0.01 & 6 & 79 & 52 & 5 \\
NGC 1778 & C0504+369 & 168.90 & $-$2.02 & 1560 & 105 & 5 & 0.32 & 0.04 & 7 & 129 & 29 & 7 \\
NGC 2192 & C0611+398 & 173.41 & +10.65 & 3467 & 72 & 4 & 0.21 & 0.01 & 4 & 1072 & 48 & 4 \\
Berkeley 69 & C0521+326 & 174.43 & $-$1.79 & 3121 & 311 & 4 & 0.67 & 0.05 & 4 & 867 & 48 & 4 \\
NGC 2281 & C0645+411 & 174.90 & +16.88 & 488 & 48 & 4 & 0.08 & 0.02 & 6 & 431 & 105 & 6 \\
Berkeley 17 & C0517+305 & 175.65 & $-$3.65 & 2575 & 88 & 4 & 0.63 & 0.05 & 4 & 9448 & 1776 & 4 \\
Melotte 25 & Hyades & 180.06 & $-$22.34 & 42 & 3 & 5 & 0.00 & 0.01 & 6 & 708 & 136 & 6 \\
NGC 1647 & C0443+189 & 180.34 & $-$16.77 & 492 & 69 & 3 & 0.36 & 0.03 & 5 & 130 & 25 & 5 \\
NGC 2158 & C0604+241 & 186.63 & +1.78 & 4403 & 589 & 5 & 0.44 & 0.06 & 5 & 1710 & 424 & 5 \\
NGC 2266 & C0640+270 & 187.79 & +10.29 & 3490 & 180 & 4 & 0.10 & 0.01 & 4 & 736 & 77 & 4 \\
NGC 2355 & C0714+138 & 203.39 & +11.80 & 2086 & 163 & 5 & 0.14 & 0.06 & 5 & 833 & 137 & 5 \\
NGC 2632 & Praesepe & 205.92 & +32.48 & 171 & 12 & 6 & 0.00 & 0.01 & 8 & 753 & 201 & 8 \\
Berkeley 32 & C0655+065 & 207.95 & +4.40 & 3491 & 401 & 5 & 0.15 & 0.03 & 5 & 3477 & 698 & 5 \\
NGC 2682 & C0847+120 & 215.70 & +31.90 & 820 & 47 & 12 & 0.05 & 0.02 & 13 & 4093 & 958 & 12 \\
Melotte 111 & Coma Ber & 221.35 & +84.03 & 86 & 7 & 5 & 0.00 & 0.01 & 6 & 522 & 82 & 6 \\
Berkeley 39 & C0744$-$044 & 223.46 & +10.10 & 4283 & 425 & 4 & 0.12 & 0.01 & 4 & 7331 & 906 & 4 \\
NGC 2548 & C0811$-$056 & 227.87 & +15.39 & 727 & 57 & 6 & 0.04 & 0.02 & 8 & 364 & 102 & 8 \\
NGC 2506 & C0757$-$106 & 230.56 & +9.94 & 3315 & 219 & 8 & 0.06 & 0.04 & 9 & 1648 & 485 & 9 \\
NGC 2539 & C0808$-$126 & 233.71 & +11.11 & 1270 & 162 & 5 & 0.08 & 0.02 & 7 & 490 & 129 & 6 \\
NGC 2527 & C0803$-$280 & 246.09 & +1.86 & 581 & 29 & 4 & 0.07 & 0.03 & 5 & 619 & 163 & 5 \\
NGC 2477 & C0750$-$384 & 253.56 & $-$5.84 & 1227 & 166 & 9 & 0.26 & 0.08 & 10 & 875 & 238 & 10 \\
Melotte 66 & C0724$-$476 & 259.56 & $-$14.24 & 3784 & 828 & 5 & 0.16 & 0.01 & 5 & 4404 & 1145 & 5 \\
Trumpler 10 & C0846$-$423 & 262.79 & +0.67 & 422 & 34 & 6 & 0.05 & 0.01 & 7 & 37 & 9 & 7 \\
NGC 2660 & C0840$-$469 & 265.93 & $-$3.01 & 2859 & 71 & 5 & 0.37 & 0.03 & 5 & 1351 & 291 & 5 \\
IC 2395 & C0839$-$480 & 266.64 & $-$3.59 & 792 & 131 & 4 & 0.09 & 0.02 & 5 & 15 & 4 & 5 \\
NGC 3105 & C0959$-$545 & 279.92 & +0.26 & 7137 & 1710 & 6 & 1.04 & 0.06 & 6 & 21 & 3 & 5 \\
IC 2581 & C1025$-$573 & 284.59 & +0.04 & 2215 & 317 & 5 & 0.42 & 0.02 & 6 & 13 & 3 & 6 \\
Bochum 10 & C1040$-$588 & 287.02 & $-$0.31 & 2472 & 551 & 4 & 0.33 & 0.03 & 5 & 8 & 2 & 4 \\
NGC 3532 & C1104$-$584 & 289.57 & +1.35 & 492 & 8 & 5 & 0.04 & 0.01 & 6 & 262 & 46 & 5 \\
IC 2714 & C1115$-$624 & 292.40 & $-$1.80 & 1217 & 72 & 5 & 0.40 & 0.06 & 6 & 253 & 66 & 6 \\
Melotte 105 & C1117$-$632 & 292.90 & $-$2.41 & 2094 & 159 & 7 & 0.48 & 0.05 & 7 & 224 & 53 & 7 \\
Stock 14 & C1141$-$622 & 295.22 & $-$0.66 & 2439 & 326 & 4 & 0.24 & 0.02 & 5 & 10 & 2 & 5 \\
Collinder 261 & C1234$-$682 & 301.68 & $-$5.53 & 2241 & 100 & 4 & 0.28 & 0.02 & 4 & 8812 & 201 & 4 \\
ESO 096$-$SC04 &  & 305.35 & $-$3.17 & 10402 & 1898 & 5 & 0.69 & 0.06 & 5 & 752 & 59 & 5 \\
\hline
\end{tabular}
\end{center}
\end{table*}
\addtocounter{table}{-1}
\begin{table*}
\caption[]{continued.} 
\begin{center}
\begin{tabular}{llrrrrrccrrrr}
\hline
\multicolumn{1}{c}{Cluster} & \multicolumn{1}{c}{Name} &  \multicolumn{1}{c}{$l$} & \multicolumn{1}{c}{$b$} & 
\multicolumn{1}{c}{$D$} & \multicolumn{1}{c}{$\sigma(D)$} & \multicolumn{1}{c}{$No$} & \multicolumn{1}{c}{$E(B-V)$} &
\multicolumn{1}{c}{$\sigma\,E(B-V)$} & \multicolumn{1}{c}{$No$} & \multicolumn{1}{c}{$age$} & \multicolumn{1}{c}{$\sigma(age)$} & 
\multicolumn{1}{c}{$No$} \\
& & & & \multicolumn{1}{c}{[pc]} & \multicolumn{1}{c}{[pc]} & & \multicolumn{1}{c}{[mag]} & 
\multicolumn{1}{c}{[mag]} & & \multicolumn{1}{c}{[Myr]} & \multicolumn{1}{c}{[Myr]} \\
\hline
NGC 5316 & C1350$-$616 & 310.23 & +0.12 & 1050 & 228 & 4 & 0.28 & 0.07 & 5 & 166 & 33 & 5 \\
Pismis 19 & C1426$-$607 & 314.71 & $-$0.30 & 2149 & 435 & 4 & 1.46 & 0.01 & 4 & 973 & 127 & 4 \\
NGC 5662 & C1431$-$563 & 316.94 & +3.39 & 684 & 60 & 7 & 0.32 & 0.01 & 8 & 77 & 20 & 8 \\
NGC 5822 & C1501$-$541 & 321.57 & +3.59 & 787 & 75 & 6 & 0.14 & 0.06 & 7 & 913 & 246 & 7 \\
NGC 6025 & C1559$-$603 & 324.55 & $-$5.88 & 725 & 94 & 5 & 0.16 & 0.01 & 6 & 74 & 22 & 6 \\
NGC 6005 & C1551$-$572 & 325.78 & $-$2.99 & 2600 & 221 & 4 & 0.48 & 0.06 & 4 & 1100 & 204 & 4 \\
NGC 6087 & C1614$-$577 & 327.73 & $-$5.43 & 893 & 57 & 7 & 0.20 & 0.03 & 9 & 78 & 19 & 9 \\
NGC 6067 & C1609$-$540 & 329.74 & $-$2.20 & 1676 & 202 & 7 & 0.36 & 0.03 & 8 & 95 & 23 & 8 \\
Lyng\aa\, 6 & C1601$-$517 & 330.37 & +0.32 & 1792 & 272 & 5 & 1.33 & 0.04 & 5 & 35 & 9 & 4 \\
NGC 6208 & C1645$-$537 & 333.76 & $-$5.76 & 1096 & 170 & 5 & 0.19 & 0.02 & 6 & 1354 & 368 & 6 \\
NGC 6253 & C1655$-$526 & 335.46 & $-$6.25 & 1567 & 108 & 7 & 0.25 & 0.08 & 7 & 3949 & 1086 & 7 \\
NGC 6250 & C1654$-$457 & 340.68 & $-$1.92 & 936 & 56 & 5 & 0.37 & 0.02 & 6 & 22 & 5 & 5 \\
NGC 6396 & C1734$-$349 & 353.93 & $-$1.77 & 1276 & 117 & 5 & 0.95 & 0.02 & 5 & 30 & 8 & 4 \\
NGC 6405 & C1736$-$321 & 356.58 & $-$0.78 & 473 & 16 & 5 & 0.14 & 0.02 & 7 & 71 & 21 & 7 \\
\hline
\end{tabular}
\end{center}
\end{table*}

\section{Comparison with the catalogue by Dias and coworkers} \label{compd}

The current best approximation, many authors use is the list of parameters
published by Dias et al. (2002). We have used 
the open-cluster parameters of the October 2005 version
of the catalogue available at http://www.astro.iag.usp.br/$\sim$wilton/ for comparison with our results. The motivation for this comparison
is to have an estimate of the statistical error introduced
when using ``mean'' catalogue parameters instead of performing
an independent analysis of its own.

We are aware that the catalogue of Dias et al. is a 
kind of compendium of different sources from the literature. They 
do not perform any weighting or averaging in the tradition of Lyng\aa.
Our analysis, on the other hand, is based on averaged values from
widely different methods and authors. It is therefore very interesting
to examine if an independent check of this catalogue can confirm its validity.

For the analysis we have calculated the differences of the
open-cluster parameters of the catalogue by Dias et al. (2002)
and our mean estimate for the sample of 395 open clusters.
For the reddening we have just used the differences as they were
calculated. Otherwise,
those difference were normalized in units of the
standard deviation of the mean for the distance and age. 
The interpretation of those parameters is the following: for values
between $-$1 and +1, the error of averaging the literature
parameter is statistically at the same level as the uncertainty 
introduced by using those catalogued ones. Fitting a Gaussian distribution
to the complete sample of values should result in a width of less than
2$\sqrt{2\ln{2}}\,\approx\,2.355$ if both ``error sources'' have the same statistical 
significance and can therefore not be disentangled (Rees 1987).

Figure \ref{dias} shows the distribution of the differences of
the distance, reddening and age as well as the basic parameters
of a Gaussian fit to these distributions. The widths of all three
distributions are in the expected range taking into account the
overall statistical errors on the cluster
parameters (Fig. \ref{means_d}). We therefore conclude that
if one uses the parameters of the catalogue by Dias et al. 
(2002) the expected errors are comparable to those which we
derived for averaging the independent values from the literature, a result which is in favor of the parameters
from the mentioned catalogue from a statistical point of view. 

\begin{figure}
\epsfxsize = 82mm
\epsffile{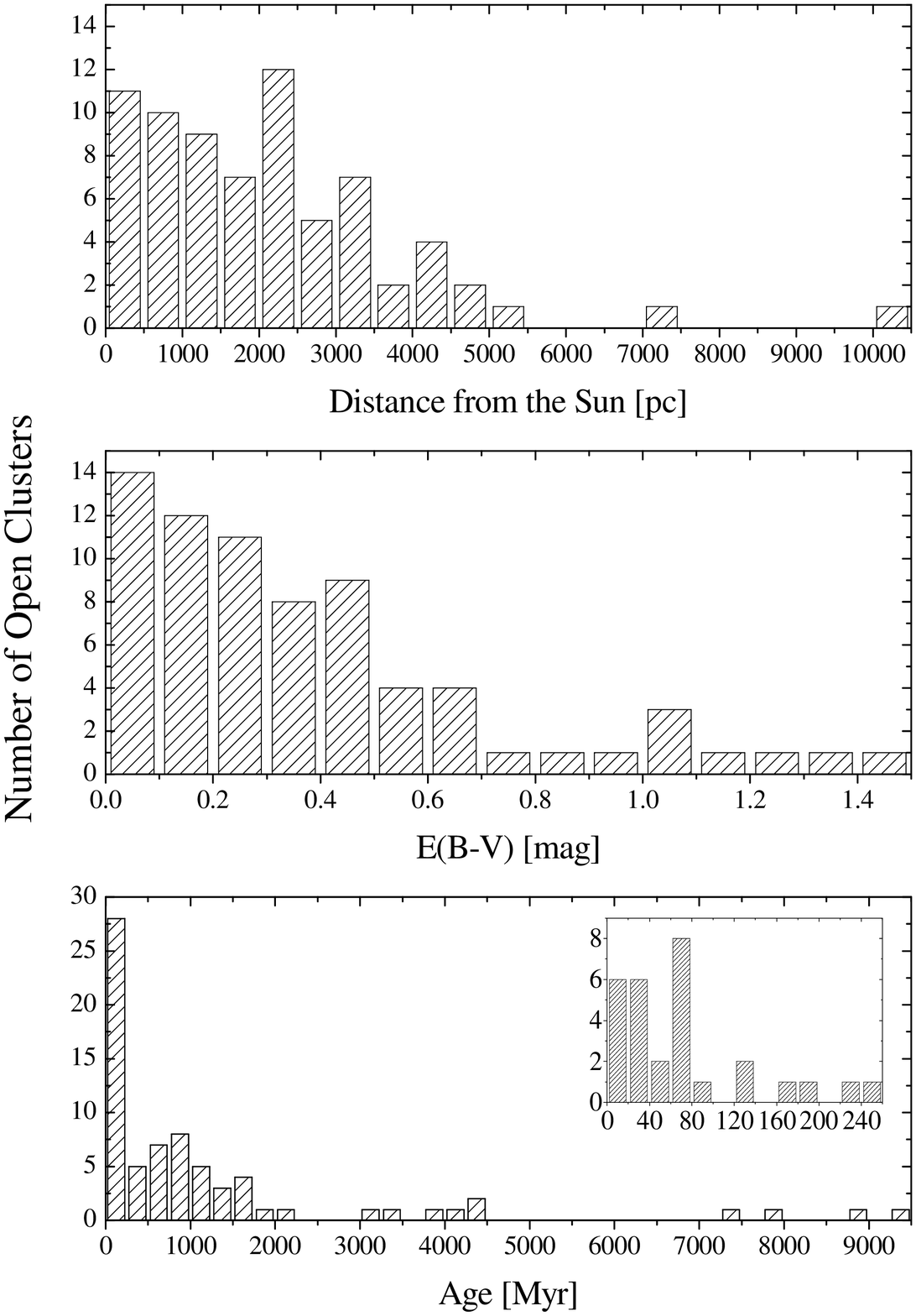}
\caption{Distribution of the distance, reddening and age of the
72 suggested standard clusters as listed in Table \ref{st_ocls}.} 
\label{st_ocls_p}
\end{figure}
 
\section{A set of standard clusters} \label{standards}

If one aims at a statistical study of a group of clusters, a
standard set for comparison is necessary. Mermilliod (1981),
for example used the following clusters to establish empirical 
isochronous curves: Alpha~Persei, Hyades, IC~4665, NGC~457,
NGC~884, NGC~2281, NGC~2287, NGC~2362, NGC~2516, NGC~3532,
NGC~3766, NGC~6231, NGC~6475 and the Pleiades. Those clusters
still serve as references when studying red giants (Eigenbrod et al. 2004).
Another interesting work in this respect is the one by Piatti,
Claria \& Bica (1998) who presented composite $M_{V}$ versus 
$(V-I)_0$ diagrams for the following ``standard'' open clusters: IC~4651,
NGC~3680, NGC~3766, NGC~6025, NGC~6067, NGC~6231, NGC~6242, NGC~6259, NGC~6451
and NGC~6633. They selected these clusters
because of accurately determined fundamental parameters. The final diagrams
are used as a homogeneous set of empirical isochrones in 
the age range between 5\,Myr and 4\,Gyr for the study of faint reddened open clusters.
The lists of Mermilliod (1981) and Piatti et al. (1998) only
coincide in three cases: NGC~3766, NGC~6231 and NGC~6475. This already
indicates that a common set of standard open clusters is very much needed.

The lists of standard clusters given in the above mentioned references 
have been chosen based on the available photometric, spectroscopic and 
kinematical data. However, as an example, Paunzen \& Maitzen (2002) have shown
that the photometric data of NGC~6451 are not intrinsically consistent and suffer
from an offset of one magnitude in Johnson $V$. Our approach is different and not 
based on the current available data for individual open clusters. On the contrary,
our list should trigger new observations to further improve the knowledge of the
listed cluster parameters. We present a list of open clusters with a significant
number of independent investigations and parameters with small errors.

We have chosen the rather strict criteria (especially for the age
estimate) for the errors of the parameters:
\begin{itemize}
\item $\sigma(D)/D$\,$\leq$\,25\,\%
\item $\sigma\,E(B-V)$\,$\leq$\,0.1\,mag
\item $\sigma(age)/age$\,$\leq$\,30\,\%
\end{itemize}
In total, 72 open clusters, listed in Table \ref{st_ocls}, fulfill these criteria.
Figure \ref{st_ocls_p} shows the distribution of the distances, reddenings and ages 
for these proposed standard clusters. They cover the complete range of known
cluster parameters within the Milky Way. These open clusters should serve as  
observational test cases for further theoretical developments. A closer inspection
of these open clusters in WEBDA reveals that, for example, the data of NGC~6208 and
NGC~6811 are still very unsatisfactory which calls for further high precision observational
investigations.

\section{Conclusions}

Using open clusters for studying global and local phenomena is very
much dependent on the accuracy of the used cluster parameters such as the
age, reddening and distance. This is also vital for testing theoretical
models. We have made a statistical study of the published cluster parameters
summarizing the results for 395 open clusters. While the reddening seems
to be quite accurate, the ages and distances suffer from severe uncertainties.
For about 90\% of all studied clusters, the error of the reddening is below 0.1\,mag.
For the distance we find absolute errors of less than 20\% for about 80\% of the
aggregates. However, only 11\% of the investigated open clusters have errors of the ages which are
less than 20\%; there are extreme discrepancies of more than 200\%. This calls for 
a homogeneous set of isochrones together with a solid fitting technique.

We have compared the errors of the mean cluster parameters with the differences
of those listed in the open cluster catalogue by Dias et al. (2002) which is
widely used. Both error distributions have the same statistical significance.
Using the parameters from the mentioned catalogue, therefore, introduces the same
error level as using averaged values from the literature, an important result 
when using this catalogue. 

As last step, a set of 72 open clusters with the most accurate (in a statistical
sense) parameters were established. Those clusters cover a wide range of reddenings,
ages as well as distances and should serve in the future as a standard set
for further observational efforts and for testing theoretical models. 
\\
\\
{\noindent \footnotesize {\bf Acknowledgements.}
We thank our referee, Jean-Claude Mermilliod, for very useful
comments which helped to improve this manuscript significantly.
This research was performed within the projects  
{\it P17580} and {\it P17920} of the Austrian Fonds zur F{\"o}rder\-ung der 
wissen\-schaft\-lichen
Forschung (FwF). M.~Netopil acknowledges 
the support by a ``Forschungsstipendium'' of the University of Vienna.
Use was made of the SIMBAD database, operated at the CDS, Strasbourg, 
the WEBDA database, operated at the University
of Vienna and the NASA Astrophysics Data System.
}

\end{document}